\documentclass[times, 10pt]{article} 
\usepackage{latex8}
\usepackage{times}

\usepackage{dsfont}

\usepackage{xypic}
\usepackage{subfigure}
\usepackage{amsthm}\theoremstyle{proof}
\usepackage{mathrsfs}
\usepackage{amsmath,amssymb}
\usepackage{xspace}
\usepackage{stmaryrd}
\usepackage{url}

\usepackage{xy}
\xyoption{frame}
\xyoption{arrow}
\xyoption{curve}
\xyoption{arc}
\xyoption{line}
\xyoption{matrix}
\xyoption{tile}
\xyoption{ps}

\makeatletter
\newdimen\inferLineSkip         \inferLineSkip=2pt
\newdimen\inferLabelSkip        \inferLabelSkip=5pt
\def\inferTabSkip{\quad}
\newdimen\@LeftOffset   % global
\newdimen\@RightOffset  % global
\newdimen\@SavedLeftOffset      % safe from users
\newdimen\UpperWidth
\newdimen\LowerWidth
\newdimen\LowerHeight
\newdimen\UpperLeftOffset
\newdimen\UpperRightOffset
\newdimen\UpperCenter
\newdimen\LowerCenter
\newdimen\UpperAdjust
\newdimen\RuleAdjust
\newdimen\LowerAdjust
\newdimen\RuleWidth
\newdimen\HLabelAdjust
\newdimen\VLabelAdjust
\newdimen\WidthAdjust
\newbox\@UpperPart
\newbox\@LowerPart
\newbox\@LabelPart
\newbox\ResultBox
\newif\if@inferRule     % whether \@infer draws a rule.
\newif\if@DoubleRule    % whether \@infer draws doulbe rules.
\newif\if@ReturnLeftOffset      % whether \@infer returns \@LeftOffset.
\newif\if@MathSaved     % whether inner math mode where \infer or
\def\DeduceSym{\vtop{\baselineskip4\p@ \lineskiplimit\z@
    \vbox{\hbox{.}\hbox{.}\hbox{.}}\hbox{.}}}
\def\@SaveMath{\@MathSavedfalse \ifmmode \ifinner
       \relax $\relax \@MathSavedtrue \fi\fi }
\def\@RestoreMath{\if@MathSaved \relax $\relax\fi }
\def\@IFnextchar#1#2#3{%
  \let\reserved@e=#1\def\reserved@a{#2}\def\reserved@b{#3}\futurelet
    \reserved@c\@IFnch}
\def\@IFnch{\ifx \reserved@c \@sptoken \let\reserved@d\@xifnch
      \else \ifx \reserved@c \reserved@e\let\reserved@d\reserved@a\else
          \let\reserved@d\reserved@b\fi
      \fi \reserved@d}
\def\@ifEmpty#1#2#3{\def\@tempa{\@empty}\def\@tempb{#1}\relax
       \ifx \@tempa \@tempb #2\else #3\fi }
\def\infer{\@SaveMath \@IFnextchar *{\@inferSteps}{\relax
       \@IFnextchar ={\@inferDoubleRule}{\@inferOneStep}}}
\def\@inferOneStep{\@inferRuletrue \@DoubleRulefalse
       \@IFnextchar [{\@infer}{\@infer[\@empty]}}
\def\@inferDoubleRule={\@inferRuletrue \@DoubleRuletrue
       \@IFnextchar [{\@infer}{\@infer[\@empty]}}
\def\@inferSteps*{\@IFnextchar [{\@@inferSteps}{\@@inferSteps[\@empty]}}
\def\@@inferSteps[#1]{\@deduce{#1}[\DeduceSym]}
\def\deduce{\@SaveMath \@IFnextchar [{\@deduce{\@empty}}
       {\@inferRulefalse \@infer[\@empty]}}
\def\@deduce#1[#2]#3#4{\@inferRulefalse
       \@infer[\@empty]{#3}{\@SaveMath \@infer[{#1}]{#2}{#4}}}
\def\@infer[#1]#2#3{\relax
       \if@ReturnLeftOffset \else \@SavedLeftOffset=\@LeftOffset \fi
       \setbox\@LabelPart=\hbox{$#1$}\relax
       \setbox\@LowerPart=\hbox{$#2$}\relax
       \global\@LeftOffset=0pt
       \setbox\@UpperPart=\vbox{\tabskip=0pt \halign{\relax
               \global\@RightOffset=0pt \@ReturnLeftOffsettrue $##$&&
               \inferTabSkip
               \global\@RightOffset=0pt \@ReturnLeftOffsetfalse $##$\cr
               #3\cr}}\relax
       \UpperLeftOffset=\@LeftOffset
       \UpperRightOffset=\@RightOffset
       \LowerWidth=\wd\@LowerPart
       \LowerHeight=\ht\@LowerPart
       \LowerCenter=0.5\LowerWidth
       \UpperWidth=\wd\@UpperPart \advance\UpperWidth by -\UpperLeftOffset
       \advance\UpperWidth by -\UpperRightOffset
       \UpperCenter=\UpperLeftOffset
       \advance\UpperCenter by 0.5\UpperWidth
       \ifdim \UpperWidth > \LowerWidth
       \UpperAdjust=0pt
       \RuleAdjust=\UpperLeftOffset
       \LowerAdjust=\UpperCenter \advance\LowerAdjust by -\LowerCenter
       \RuleWidth=\UpperWidth
       \global\@LeftOffset=\LowerAdjust
       \else   % \UpperWidth <= \LowerWidth
       \ifdim \UpperCenter > \LowerCenter
       \UpperAdjust=0pt
       \RuleAdjust=\UpperCenter \advance\RuleAdjust by -\LowerCenter
       \LowerAdjust=\RuleAdjust
       \RuleWidth=\LowerWidth
       \global\@LeftOffset=\LowerAdjust
       \else   % \UpperWidth <= \LowerWidth
       \UpperAdjust=\LowerCenter \advance\UpperAdjust by -\UpperCenter
       \RuleAdjust=0pt
       \LowerAdjust=0pt
       \RuleWidth=\LowerWidth
       \global\@LeftOffset=0pt
       \fi\fi
       \if@inferRule
       \setbox\ResultBox=\vbox{
               \moveright \UpperAdjust \box\@UpperPart
               \nointerlineskip \kern\inferLineSkip
               \if@DoubleRule
               \moveright \RuleAdjust \vbox{\hrule width\RuleWidth
                       \kern 1pt\hrule width\RuleWidth}\relax
               \else
               \moveright \RuleAdjust \vbox{\hrule width\RuleWidth}\relax
               \fi
               \nointerlineskip \kern\inferLineSkip
               \moveright \LowerAdjust \box\@LowerPart }\relax
       \@ifEmpty{#1}{}{\relax
       \HLabelAdjust=\wd\ResultBox     \advance\HLabelAdjust by -\RuleAdjust
       \advance\HLabelAdjust by -\RuleWidth
       \WidthAdjust=\HLabelAdjust
       \advance\WidthAdjust by -\inferLabelSkip
       \advance\WidthAdjust by -\wd\@LabelPart
       \ifdim \WidthAdjust < 0pt \WidthAdjust=0pt \fi
       \VLabelAdjust=\dp\@LabelPart
       \advance\VLabelAdjust by -\ht\@LabelPart
       \VLabelAdjust=0.5\VLabelAdjust  \advance\VLabelAdjust by \LowerHeight
       \advance\VLabelAdjust by \inferLineSkip
       \setbox\ResultBox=\hbox{\box\ResultBox
               \kern -\HLabelAdjust \kern\inferLabelSkip
               \raise\VLabelAdjust \box\@LabelPart \kern\WidthAdjust}\relax
       }\relax % end @ifEmpty
       \else % \@inferRulefalse
       \setbox\ResultBox=\vbox{
               \moveright \UpperAdjust \box\@UpperPart
               \nointerlineskip \kern\inferLineSkip
               \moveright \LowerAdjust \hbox{\unhbox\@LowerPart
                       \@ifEmpty{#1}{}{\relax
                       \kern\inferLabelSkip \unhbox\@LabelPart}}}\relax
       \fi
       \global\@RightOffset=\wd\ResultBox
       \global\advance\@RightOffset by -\@LeftOffset
       \global\advance\@RightOffset by -\LowerWidth
       \if@ReturnLeftOffset \else \global\@LeftOffset=\@SavedLeftOffset \fi
       \box\ResultBox
       \@RestoreMath
}
\makeatother

\def\makenewenum#1#2{%
\newcounter{cnt#1}
\newenvironment{#1}%
{\begin{list}{\makebox[0pt][r]{#2}}%
{\setlength{\itemsep}{0pt}%
 \setlength{\parsep}{.2em}%
 \setlength{\leftmargin}{2.5em}%
 \setlength{\labelwidth}{.2em}%
 \usecounter{cnt#1}}}%
{\end{list}}}
\makenewenum{enu}{\arabic{cntenu}.}
\makenewenum{enum}{\em(\arabic{cntenum})}
\makenewenum{itenum}{\em(\arabic{cntenum})}
\makenewenum{en}{(\roman{cnten})}
\makenewenum{renum}{\em(\roman{cntrenum})}
\makenewenum{enhyphen}{(\roman{cntenhyphen}')}

\newcommand{\varimp}{\supset} 

\newcommand{\MSQR}{\textsf{MSQR}} 
\newcommand{\MSpQR}{\textsf{MSpQR}}

\newcommand{\measB}{\blacksquare}
\newcommand{\measD}{\blacklozenge}
\newcommand{\uniB}{\square}
\newcommand{\uniD}{\lozenge}
\newcommand{\Un}{\textsf{U}}
\newcommand{\Me}{\textsf{M}}

\newtheorem{theorem}{Theorem}

\newtheorem{lemma}{Lemma}
\newtheorem{definition}{Definition}
\newtheorem{fact}{Fact}

\mathcode`\:="003A
\newcommand{\RAA}{\mathit{RAA}}

\newcommand{\Unr}{\Un \mathit{refl}}
\newcommand{\Uns}{\Un \mathit{symm}}
\newcommand{\Unt}{\Un \mathit{trans}}

\renewcommand{\Un}{\mathsf{U}}
\renewcommand{\Me}{\mathsf{M}}

\newcommand{\PMe}{\mathsf{P}}

\newcommand{\class}{\mathit{class}}
\newcommand{\Meser}{\Me\mathit{ser}}
\newcommand{\Mesrefl}{\Me\mathit{srefl}}
\newcommand{\Metrans}{\Me\mathit{trans}}
\newcommand{\PMetrans}{\PMe\mathit{trans}}
\newcommand{\Mesubone}{\Me\mathit{sub1}}
\newcommand{\Mesubtwo}{\Me\mathit{sub2}}
\newcommand{\PMesubone}{\PMe\mathit{sub1}}
\newcommand{\PMesubtwo}{\PMe\mathit{sub2}}

\newcommand{\LF}{\mathit{LF}}
\newcommand{\RF}{\mathit{RF}}

\newcommand{\I}{\mathscr{I}}

\makeatletter
\DeclareSymbolFont{symbolsC}{U}{pxsyc}{m}{n}
\SetSymbolFont{symbolsC}{bold}{U}{pxsyc}{bx}{n}
\DeclareFontSubstitution{U}{pxsyc}{m}{n}
\def\re@DeclareMathSymbol#1#2#3#4{%
    \let#1=\undefined
    \DeclareMathSymbol{#1}{#2}{#3}{#4}}
    \re@DeclareMathSymbol{\Diamonddot}{\mathord}{symbolsC}{144}
\makeatother

\title{A Qualitative Modal Representation of Quantum Register Transformations \\
(Extended Version)}
\author{Andrea Masini \qquad Luca Vigan\`o \qquad Margherita Zorzi  \\
Department of Computer Science \\
University of Verona, Italy \\
\{andrea.masini $\mid$ luca.vigano $\mid$ margherita.zorzi\}@univr.it}

\begin{document}

\maketitle
\thispagestyle{empty}

\begin{abstract}
  We introduce two modal natural deduction systems that are suitable to
  represent and reason about transformations of quantum registers in an abstract, qualitative, way.
  Quantum registers represent quantum systems, and can be viewed as the 
  structure of quantum data for quantum operations. 
  Our systems provide a modal framework for reasoning about operations on quantum registers (unitary transformations and measurements)
  in terms of possible worlds (as abstractions of quantum registers) and accessibility relations between these worlds. We give a Kripke--style
  semantics that formally describes quantum register transformations, and prove the 
  soundness and completeness of our systems with respect to this semantics.
\end{abstract}

\Section{Introduction}
Quantum computing 
defines an alternative computational paradigm, based on a quantum 
model~\cite{BasDa05short} rather than a classical one. The basic units of the quantum model are the 
\textit{quantum bits}, or \textit{qubits} for short (mathematically, normalized vectors of the Hilbert Space 
$\mathds{C}^{2}$).  Qubits represent informational units and can assume both classical values 0 and 1, and all their superpositional values.

A \emph{quantum register} is a generalization of the qubit: a generic quantum register is the representation of a quantum state of $n$ qubits (mathematically, it is a normalized vector of the
Hilbert space $\mathds{C}^{2^n}$). In this paper, we are not interested in the structure of  quantum registers, but rather in the way quantum registers are transformed. Hence, we will abstract away from the internals of quantum registers and represent them in a generic way in order to 
describe how operations transform a register into another one.

It is possible to modify a quantum register in two ways: by applying a unitary transformation 
or by measuring.  Unitary transformations (corresponding to the so-called unitary operators of 
the Hilbert space) model the internal evolution of a quantum system, whereas measurements 
correspond to the results of the interaction between a quantum system and an observer.  
The outcome of an observation can be either the reduction to a quantum
state or the reduction to a classical (non quantum) state. 
In particular, in this paper, we say that a quantum register $w$ is \emph{classical} iff $w$ is idempotent with respect to measurement, i.e.~each measurement of $w$ has $w$ as outcome. 
We call a measurement \emph{total} when the outcome of the measurement 
is a classical register.

We propose to model measurement and unitary transformations by means of suitable modal 
operators. More specifically, the main contribution of this paper is the formalization of 
a modal natural deduction system~\cite{Prawitz65,TroelstraSchwichtenberg96} 
in order to represent (in an abstract, qualitative, way) the fundamental operations on quantum registers: unitary transformations and total measurements. We call this system $\MSQR$. We also formalize a variant of this system, called $\MSpQR$, to represent the case of generic (not necessarily total) measurements. 

It is important to observe that our logical systems are not a quantum
logic. Since 1936~\cite{vN36}, various logics have been investigated
as a means to formalize reasoning about propositions taking into account
the principles of quantum theory,
e.g.~\cite{DallaCh77,DallaChiara86}. In general, it is
possible to view quantum logic as a logical axiomatization of quantum
theory, which provides an adequate foundation for a theory of
reversible quantum processes,
e.g.~\cite{AbDu06,BaltSmets04,BalSmets06,Mittel79}.

Our work moves from quite a different point of view: we do not aim to propose a general logical formalization of quantum theory, rather we describe how it is possible to use modal logic to
reason in a simple way about quantum register transformations.  
Informally, in our proposal, a modal world represents (an abstraction of) a quantum register.
The discrete temporal evolution of a quantum register is controlled and determined by a 
sequence of unitary transformations and measurements that can change the description of a
quantum state into other descriptions. So, the evolution of a quantum register can be viewed as 
a graph, where the nodes are the (abstract) quantum registers and the arrows represent quantum transformations. The arrows give us the so-called accessibility relations of Kripke models and two
nodes linked by an arrow represent two related quantum states: the target node is obtained from
the source node by means of the operation specified in the decoration of the arrow.

Modal logic, as a logic of possible worlds, is thus a natural way to
represent this description of a quantum system: the worlds model the
quantum registers and the relations of accessibility between worlds
model the dinamical behavior of the system, as a consequence of the
application of measurements and unitary transformations. To emphasize
this semantic view of modal logic, we give our deduction system in the
style of \emph{labelled
  deduction}~\cite{Gabbay96,Simpson93,Vigano00a}, a framework for
giving uniform presentations of different non-classical logics.  The
intuition behind labelled deduction is that the labelling (sometimes
also called prefixing, annotating or subscripting) allows one to
explicitly encode in the syntax additional information, of a semantic
or proof-theoretical nature, that is otherwise implicit in the logic
one wants to capture. Most notably, in the case of modal logic, this
additional information comes from the underlying Kripke semantics: the
labelled formula $x:A$ intuitively means that $A$ holds at the world
denoted by the label $x$ within the underlying Kripke structure
(i.e.~model), and labels also allow one to specify at the syntactic
level how the different worlds are related in the Kripke structures
(e.g.~the formula $x R y$ specifies that the world denoted by $y$ is
accessible from that denoted by $x$).

We proceed as follows. In Section~\ref{sec:syntax}, we define the labelled modal natural deduction system $\MSQR$, which contains two modal operators suitable to represent and reason about unitary transformations and total measurements of quantum registers. In Section~\ref{sec:semantics}, we 
give a  possible worlds semantics that formally describes these quantum register transformations, 
and prove the soundness and completeness of $\MSQR$ with respect to this semantics. In Section~\ref{sec:partial}, we formalize $\MSpQR$, a variant of $\MSQR$ that provides a modal system 
representing all the possible (thus not necessarily total) measurements.  We
conclude in Section~\ref{sec:conclusions} with a brief summary and a discussion of future work.
Full proofs of the technical results are given in the appendix.

\section{The deduction system $\MSQR$}
\label{sec:syntax}

Our labelled modal natural deduction system $\MSQR$, which formally represents unitary transformations and total measurements of quantum registers,
comprises of rules that derive formulas of two kinds: modal formulas and relational formulas. We 
thus define a modal language and a relational language. 

The alphabet of the \emph{relational language} consists of:
\begin{itemize}
\item the binary symbols $\Un$ and $\Me$,
\item a denumerable set $x_0, x_1, \ldots$ of \textit{labels}.
\end{itemize}
Metavariables $x,y,z$, possibly annotated with subscripts and superscripts, range over the set of 
labels.
For brevity, we will sometimes speak of a ``world" $x$ meaning that the label $x$ stands for a world
$\I(x)$, where $\I$ is an interpretation function mapping labels into 
worlds as formalized in Definition~\ref{def:semantics} below.

The set of \textit{relational formulas} (\textit{r--formulas}) is given
by expressions of the form $x\Un y$ and $x\Me y$.

The alphabet of the \textit{modal language}  consists of:
\begin{itemize}
\item a denumerable set $r, r_0, r_1,\ldots$  of \textit{propositional symbols},
\item the standard \textit{propositional connectives} $\bot$ and $\varimp$,
\item the unary \textit{modal operators} $\uniB$ and $\measB$.
\end{itemize}
The set of \emph{modal formulas} (\emph{m--formulas}) is the least
set that contains $\bot$ and the propositional symbols, and is closed
under the propositional connectives and the modal operators.  
Metavariables $A$, $B$, $C$, possibly indexed, range over modal formulas.
Other connectives can be defined in the usual manner, e.g.~$\neg A \equiv A\varimp \bot$, 
$A \wedge B \equiv \neg (A \varimp \neg B)$, 
$A \leftrightarrow B \equiv (A \varimp B) \wedge (B \varimp A)$,
$\Diamond A \equiv \neg \Box \neg A$, $\measD A \equiv \neg \measB \neg A$, etc.

Let us give, in a rather informal way, the intuitive meaning of the modal operators of our language:
\begin{itemize}
 \item $\uniB A$ means: $A$ is true after the application of any
   unitary transformation.
\item $\measB A$ means: $A$ is true in each quantum register obtained by a total
  measurement.
\end{itemize}

A \emph{labelled formula} (\emph{l--formula}) 
is an expression $x:A$, where $x$ is a label and $A$ is an m--formula. 
A \emph{formula} is either an r--formula or an l--formula. The metavariable $\alpha$, possibly 
indexed, ranges over formulas. We write $\alpha(x)$ to denote that the label $x$ occurs in the 
formula $\alpha$, so that $\alpha(y/x)$ denotes the substitution of the label $y$ for all occurences of 
$x$ in $\alpha$.

\begin{figure*}[t]
  \begin{displaymath}
    \renewcommand{\arraystretch}{4}
    \begin{array}{c}
    \infer[\varimp I]{x: A\varimp B}{\infer*{x:B}{[x:A]}}
    \qquad
    \infer[\varimp E]{x: B}{x: A \varimp B & x:B}
    \qquad
    \infer[\mathit{RAA}]{x: A}{\infer*{y:\bot}{[x:\neg A]}}
    \qquad
    \infer[\bot E]{\alpha}{x:\bot}
    \qquad
    \infer[\bigstar I^*]{x: \bigstar A}{\infer*{y: A}{[x R y]}}
    \qquad
    \infer[\bigstar E]{y: A}{x: \bigstar A & x R y}
    \\
    \infer[\Unr]{x \Un x}{}
   \qquad
    \infer[\Uns]{y \Un x}{x \Un y}
    \qquad
    \infer[\Unt]{x \Un z}{x \Un y & y \Un z}
    \qquad 
    \infer[\Un I]{x \Un y}{x \Me y}
    \qquad
    \infer[\Meser^*]{\alpha}{\infer*{\alpha}{[x \Me y]}}
    \qquad
    \infer[\Mesrefl]{y \Me y}{x \Me y}
    \\
    \infer[\Mesubone]{\alpha(y/x)}{\alpha(x) & x \Me x & x \Me y}
    \qquad
    \infer[\Mesubtwo]{\alpha(x/y)}{\alpha(y) & x \Me x & x \Me y}
    \end{array}
  \end{displaymath}
  In $\bigstar I$, $y$ is fresh: it is different from $x$ and does not occur in any assumption on which $y:A$ depends other than $x R y$. \\
  In $\Meser$, $y$ is fresh: it is different from $x$ and does not occur in $\alpha$ nor in any assumption on which $\alpha$ depends other than $x \Me y$.
  \caption{The rules of $\MSQR$}
  \label{fig:rules}
\end{figure*}

Figure~\ref{fig:rules} shows the rules of $\MSQR$, where the notion of \emph{discharged/open assumption} is standard~\cite{Prawitz65,TroelstraSchwichtenberg96}, e.g.~the formula $[x:A]$ is 
discharged in the rule $\varimp I$:
\begin{description}
\item[Propositional rules:] The rules $\varimp I$, $\varimp E$ and $\RAA$ are just the labelled version of the standard (\cite{Prawitz65,TroelstraSchwichtenberg96}) natural deduction rules for implication introduction and elimination and for 
  \emph{reductio ad absurdum}, where we do not enforce Prawitz's side
  condition that $A \neq \bot$.\footnote{See~\cite{Vigano00a} for a
    detailed discussion on the rule $\RAA$, which in particular
    explains how, in order to maintain the duality of modal operators
    like $\Box$ and $\Diamond$, the rule must allow one to derive
    $x:A$ from a contradiction $\bot$ at a possibly different world
    $y$, and thereby discharge the assumption $x: \neg A$.} The
  ``mixed'' rule $\bot E$ allows us to derive a generic formula
  $\alpha$ whenever we have obtained a contradiction $\bot$ at a world
  $x$.
\item[Modal rules:] We give the rules for a generic modal operator $\bigstar$, with a corresponding generic accessibility relation $R$, since all the modal operators share the structure of these basic introduction/elimination rules; this holds because, for instance, we express $x: \Box A$ as the metalevel 
implication $x \Un y \Longrightarrow y:A$ for an arbitrary $y$ accessible from $x$. In particular:
\begin{itemize}
\item if  $\bigstar$ is $\uniB$ then $R$ is $\Un$,
\item if $\bigstar$ is $\measB$ then $R$ is $\Me$.
\end{itemize}
\item[Other rules:] \
\begin{itemize} 
\item In order to axiomatize $\uniB$, 
we add rules $\Unr$, $\Uns$, and $\Unt$, formalizing that $\Un$ is an equivalence relation.
\item 
In order to axiomatize $\measB$, 
we add rules formalizing the following properties:
\begin{itemize}
\item If $x\Me y$ then there is specific unitary transformation
  (depending on $x$ and $y$) that generates $y$ from $x$: rule $\Un I$.
\item The total measurement process is serial: rule $\Meser$ says that if from the assumption $x \Me y$ we can derive $\alpha$ for a \emph{fresh} $y$ (i.e.~$y$ is different from $x$ and does not occur in $\alpha$ nor in any assumption on which $\alpha$ depends other than $x \Me y$), then we can discharge the assumption (since there always is some $y$ such that $x \Me y$) and conclude $\alpha$.
\item The total measurement process is shift-reflexive: rule $\Mesrefl$.
\item Invariance with respect to classical worlds: rules $\Mesubone$ and $\Mesubtwo$ say that, if 
$x \Me x$ and $x \Me y$, then $y$ must be equal to $x$ and so we can substitute the one for the other in any formula $\alpha$.
\end{itemize}
\end{itemize}
\end{description}

\begin{definition}[Derivations and proofs]
  A \emph{derivation} of a formula $\alpha$ from a set of formulas
  $\Gamma$ in $\MSQR$ is a tree formed using the rules in $\MSQR$,
  ending with $\alpha$ and depending only on a finite subset of
  $\Gamma$; we then write $\Gamma \vdash \alpha$.  A derivation of
  $\alpha$ in $\MSQR$ depending on the empty set, $\vdash \alpha$, is
  a \emph{proof} of $\alpha$ in $\MSQR$ and we then say that $\alpha$
  is a theorem of $\MSQR$.
\end{definition}

For instance, the following labelled formula schemata are all provable
in $\MSQR$ (where, in parentheses, we give the intuitive meaning of
each formula in terms of quantum register transformations):
\begin{enumerate}
\item $x: \uniB A \varimp A $\\ (the identity transformation is unitary).
\item $x: A \varimp \uniB\uniD A$ \\ (each unitary transformation is invertible).
\item $x: \uniB A \varimp \uniB\uniB A  $\\ (unitary transformations are composable).
\item $x: \measB A \varimp \measD A$ \\
(it is always possible to perform a total measurement of a quantum register).
\item\label{example-one} $x: \measB (A \leftrightarrow \measB A)$ \\ 
(it is always possible to perform a total measurement with a complete reduction of a quantum register to a classical one).
\item\label{example-two}
 $x: \measB A \varimp \measB\measB A$ \\ 
 (total measurements are composable).
\end{enumerate}

As concrete examples, Figure~\ref{fig:examples} contains the proofs of the formulas~\ref{example-one} and~\ref{example-two}, where, for simplicity, here and in the following (cf.~Figure~\ref{fig:example-MSpQR}), we employ the rules for 
equivalence ($\leftrightarrow I$) and for negation ($\neg I$ and $\neg E$), 
which are derived from the propositional rules as is standard. For instance, 
\begin{displaymath}
\begin{array}{ccc}
\vcenter{
\infer[\neg I^1]{x: \neg A}{\infer*{y: \bot}{[x:A]^1}}
}
&
\textrm{abbreviates}
&   
\vcenter{
\infer[\varimp I^1]{x: A \varimp \bot}
 {
 \infer[\bot E \textrm{ (or } \RAA \textrm{)}]{x: \bot}
  {
  \infer*{y: \bot}{[x:A]^1}
  }
 }
}
\end{array}
\end{displaymath}
We can similarly derive rules about r--formulas. For instance, we can derive a rule for the transitivity 
of $\Me$ as shown at the top of the proof of the formula~\ref{example-two} in Figure~\ref{fig:examples}:
\begin{displaymath}
\infer[\Metrans]{x \Me z}{x \Me y & y \Me z}
\end{displaymath}
abbreviates
\begin{displaymath}
\infer[\Mesubone]{x \Me z}
 {
 y \Me z
 &
 \infer[\Mesrefl]{z \Me z}{y \Me z}
 &
 x \Me y
 }
\end{displaymath}

\begin{figure*}[t]
  \begin{displaymath}
    \begin{array}{c}
    \infer[\measB I^1]{x: \measB (A \leftrightarrow \measB A)}
     {
     \infer[\leftrightarrow I]{y: A \leftrightarrow \measB A}
      {
      \infer[\varimp I^2]{y: A \varimp \measB A}
       {
       \infer[\measB I^3]{y: \measB A}
        {
        \infer[\Mesubone]{z: A}
         {
         [y:A]^2
         &
         \infer[\Mesrefl]{y \Me y}{[x \Me y]^1}
         & 
         [y \Me z]^3
         }
        }
       }
      & 
      \infer[\varimp I^4]{y: \measB A \varimp A}
       {
       \infer[\Mesubone]{y: A}
        {
        [y: \measB A]^4
        &
        \infer[\Mesrefl]{y \Me y}{[x \Me y]^1}
        }
       }
      }
     } 
     \\[2em]
     \infer[\varimp I^1]{x: \measB A \varimp \measB\measB A}
      {
      \infer[\measB I^2]{x: \measB\measB A}
       {
       \infer[\measB I^3]{y: \measB A}
        {
        \infer[\measB E]{z: A}
         {
         [x: \measB A]^1
         &
         \infer[\Mesubone]{x \Me z}
          {
          [y \Me z]^3
          &
          \infer[\Mesrefl]{z \Me z}{[y \Me z]^3}
          &
          [x \Me y]^2
          }
         }
        }
       }
      }
     \end{array}
     \end{displaymath}
  \caption{Examples of proofs in $\MSQR$}
  \label{fig:examples}
\end{figure*}

\Section{A semantics for unitary transformations and total measurements}
\label{sec:semantics}

We give a semantics that formally describes unitary transformations and total measurements of quantum registers, and then prove that 
$\MSQR$ is sound and complete with respect to this semantics. Together with the corresponding result for generic measurements in Section~\ref{sec:partial}, this means that our modal systems indeed provide a representation of quantum registers and operations on them, which was the main goal of the paper.

\begin{figure*}
\begin{displaymath}
\begin{array}{llll}
\xymatrix{
v \ar[r]^M & w &
}
&
\xymatrix{
v \ar[r]^M & w\ar@(r,u)[]_M & &
}
&
\xymatrix{
v\ar@(r,u)[]_M \ar@{-//}[r]_M \ar@{.>}[dr]|-{?}_U & \\
& & 
}
&
\xymatrix{
v \ar[r]^M & w\ar@(r,u)[]_M \ar@{-//}[r]_M \ar@{.>}[dr]|-{?}_U & \\
& & 
} \\
\text{(ii)} & \text{(iii)} & \text{(iv)} & \text{(iii) and (iv)}
\end{array}
\end{displaymath}
\caption{Some properties of the relation $M$}
\label{fig:properties}
\end{figure*}
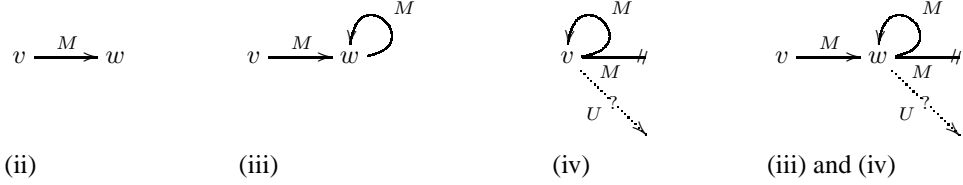

\begin{definition}[Frames, models, structures]\label{def:semantics}
  A \emph{frame} 
  is a tuple $\mathscr{F} =
  \langle W, U, M\rangle$, where:
  \begin{itemize}
  \item $W$ is a non-empty set of \emph{worlds} \\
  (representing abstractly the quantum registers);
  \item $U\subseteq W\times W$ is an equivalence relation \\
  ($v U w$ means that $w$ is obtained by applying a unitary transformation to $v$; 
  $U$ is an equivalence relation since identity is a unitary transformation, 
  each unitary transformation must be invertible, and unitary transformations are composable);
  \item $M \subseteq W \times W$ \\
  ($v M w$ means that $w$ is obtained by means of a total measurement of $v$);
  \end{itemize}
with the following properties:
\begin{renum}
\item $\forall v, w.\ v M w \Longrightarrow v U w$ 
\item   $\forall v.\ \exists w.\ v M w$ 
\item $\forall v, w.\ v M w \Longrightarrow w M w$ 
\item $\forall v, w.\ v M v \ \& \ v M w \Longrightarrow v = w$ 
\end{renum}
\emph{(i)} means that although it is not true that measurement is a unitary transformation, locally for each $v$, if $v M w$ then there is a particular unitary transformation, depending on $v$ and $w$, that generates $w$ from $v$; the vice versa cannot hold, since in quantum theory measurements cannot be used to obtain the unitary evolution of a quantum system.
\emph{(ii)} means that each quantum register is totally measurable.
\emph{(iii)} and \emph{(iv)} together mean that after a total measurement we obtain a classical world.
Figure~\ref{fig:properties} shows properties \emph{(ii)}, \emph{(iii)} and \emph{(iv)}, respectively, as well as the combination of \emph{(iii)} and \emph{(iv)}.\footnote{Note that while (iv) says that $v$ is idempotent with respect to $M$, a unitary transformation $U$ could still be applied to $v$ (and hence the dotted arrow decorated with a ``?" for $U$).}
  
A \emph{model} is a pair $\mathscr{M}=\langle\mathscr{F}, V\rangle$,
where $\mathscr{F}$ is a frame and $V\,:\, W\rightarrow
2^{\mbox{\textit{Prop}}}$ is an interpretation function mapping worlds
into sets of formulas.
  
A \emph{structure} 
is a pair $\mathscr{S}=\langle\mathscr{M}, \I\rangle$, where 
$\mathscr{M}$ is a model and $\I\, :\, \mathit{Var} \rightarrow  W$ is an interpretation function mapping 
variables (labels) into worlds in $W$, and mapping a relation symbol $R \in \{\Un,\Me\}$
into the corresponding frame relation $\I(R) \in \{U,M\}$. We extend $\I$ to 
formulas and sets of formulas in the obvious way: $\I(x:A)=\I(x):A$,
$\I(x R y) = \I(x) \I(R) \I(y)$, and $\I(\{\alpha_1,\ldots,\alpha_n\})=\{\I(\alpha_1),\ldots,\I(\alpha_n)\}$. 
\end{definition}

Given this semantics, we can define what it means for formulas to be true, and then prove
the soundness and completeness of $\MSQR$.

\begin{definition}[Truth]\label{def:truth}
  \emph{Truth} for an m--formula in a model $\mathscr{M} =
  \langle W, U, M, V\rangle$
  is the smallest relation $\vDash$ satisfying:
  \begin{center}
    \begin{tabular}{lll}
    $\mathscr{M},w \vDash r$ & iff & $r \in V(w)$ \\
    $\mathscr{M},w \vDash A \varimp B$ & iff & $\mathscr{M},w \vDash A \Longrightarrow
    \mathscr{M},w \vDash B$ \\
    $\mathscr{M},w\vDash \uniB A$ & iff & 
    $\forall w'.\  w U w' \Longrightarrow \mathscr{M},w'\vDash A$\\
    $\mathscr{M},w\vDash \measB A$ & iff & 
    $\forall w'.\ w M w' \Longrightarrow \mathscr{M}, w'\vDash A$ \\
    \end{tabular}
  \end{center}
  Thus, for an m--formula $A$, we write $\mathscr{M} \vDash A$  iff 
  $\mathscr{M},w \vDash A$ for all $w$. 

  \emph{Truth} for a formula $\alpha$ in a structure 
  $\mathscr{S}=\langle\mathscr{M}, \I\rangle$
  is then the smallest relation $\vDash$ satisfying:
  \begin{center}
    \begin{tabular}{lll}
    $\mathscr{M},\I\vDash x\Me y$ & iff & 
    $\I(x) M \I(y)$ \\
    $\mathscr{M},\I\vDash x\Un y$ & iff & 
    $\I(x) U \I(y)$ \\
    $\mathscr{M},\I\vDash x:A$ & iff & 
    $\mathscr{M}, \I(x)\vDash A$
    \end{tabular}
  \end{center}
We will omit $\mathscr{M}$ when it is not relevant, and we will denote $\I \vDash x:A$ also by 
$\vDash \I(x):A$ or even $\vDash w:A$ for $\I(x)=w$.
By extension, $\mathscr{M},\I\vDash \Gamma$ iff 
$\mathscr{M},\I\vDash \alpha$ for all $\alpha$ in the set of formulas $\Gamma$. Thus, for 
a set of formulas $\Gamma$ and a formula $\alpha$,
\begin{displaymath}
\begin{array}{lll}
\Gamma \vDash \alpha & \mathrm{iff} & 
\forall \mathscr{M}, \I.\ \mathscr{M} \vDash \I(\Gamma) \Longrightarrow \mathscr{M} 
\vDash \I(\alpha) \\
& \mathrm{iff} & 
\forall \mathscr{M}, \I.\ \mathscr{M}, \I \vDash \Gamma \Longrightarrow \mathscr{M}, \I\vDash \alpha
\end{array}
\end{displaymath}
\end{definition}

By adapting standard proofs (see, e.g., \cite{Gabbay96,Prawitz65,Simpson93,TroelstraSchwichtenberg96,Vigano00a} and the proofs in the appendix), we have:

\begin{theorem}[Soundness and completeness of $\MSQR$]\label{theorem:soundness-completeness}
$\Gamma \vdash \alpha$ iff $\Gamma \vDash \alpha$.
\end{theorem}

\Section{Generic measurements}
\label{sec:partial}

\begin{figure*}[t]
  \begin{displaymath}
    \renewcommand{\arraystretch}{4}
  \begin{array}{l}
    \varimp I, \ \varimp E,\ \mathit{RAA},\ \bot E,\ \bigstar I^*,\ \bigstar E,\ \Unr,\ \Uns,\ \Unt,\\
    \infer[\PMe\Un I]{x \Un y}{x \PMe y}
    \qquad
    \infer[\PMetrans]{x \PMe z}{x \PMe y & y \PMe z}
    \qquad
    \infer[\class^*]{\alpha}{\infer*{\alpha}{[x \PMe y]\, [y \PMe y]}}
    \qquad
    \infer[\PMesubone]{\alpha(y/x)}{\alpha(x) & x \PMe x & x \PMe y}
    \qquad
    \infer[\PMesubtwo]{\alpha(x/y)}{\alpha(y) & x \PMe x & x \PMe y}
    \end{array}
  \end{displaymath}
  In $\bigstar I$, $y$ is fresh: it is different from $x$ and does not occur in any assumption on which $y:A$ depends other than $x R y$. \\
  In $\class$, $y$ is fresh: it is different from $x$ and does not occur in $\alpha$ nor in any assumption on which $\alpha$ depends other than $x \PMe y$ and $y \PMe y$.
  \caption{The rules of $\MSpQR$}
  \label{fig:rules-MSpQR}
\end{figure*}

\begin{figure*}[t]
  \begin{displaymath}
     \begin{array}{c}
    \infer[\class^1]{x: \neg \boxdot \neg (A \varimp \boxdot A)}
     {
     \infer[\neg I^2]{x: \neg \boxdot \neg (A \varimp \boxdot A)}
      {
      \infer[\neg E]{y:\bot}
       {
       \infer[\boxdot E]{y: \neg (A \varimp \boxdot A)}
        {
        [x: \boxdot \neg (A \varimp \boxdot A)]^2 & [x \PMe y]^1
        }
       &
       \infer[\varimp I^3]{y: A \varimp \boxdot A}
        {
        \infer[\boxdot I^4]{y: \boxdot A}
         {
         \infer[\PMesubone]{z:A}
          {
          [y:A]^3 & [y \PMe y]^1& [y \PMe z]^4 
          }
         }
        }
       }
      }
     } 
    \end{array}
   \end{displaymath}
  \caption{An example proof in $\MSpQR$}
  \label{fig:example-MSpQR}
\end{figure*}

In quantum computing, not all measurements are required to be total: think, for example, of the case of observing only the first qubit of a quantum register. To this end, in this section, we formalize $\MSpQR$, a variant of $\MSQR$ that provides a modal system representing all the possible 
(thus not necessarily total) measurements. 
We obtain $\MSpQR$ from $\MSQR$ by means of the following changes: 
\begin{itemize}
\item The alphabet of the modal language contains the unary modal operator $\boxdot$ instead of $\measB$, with corresponding $\Diamonddot$, where $\boxdot A$ intuitively means that $A$ is true in each quantum register obtained by a measurement.
\item The set of relational formulas contains expressions of the form $x \PMe y$ instead of $x \Me y$.
\item The rules of $\MSpQR$ are given in Figure~\ref{fig:rules-MSpQR}. In particular, $\bigstar$ is either 
$\uniB$ (as before) or $\boxdot$, for which then $R$ is $\PMe$, and whose properties are formalized by  the following additional rules:
\begin{itemize}
\item If $x \PMe y$ then there is a specific unitary transformation
  (depending on $x$ and $y$) that generates $y$ from $x$: rule $\PMe\Un I$.
\item The measurement process is transitive: rule $\PMetrans$.
\item There are (always reachable) classical worlds: $\class$ says that $y$ is a classical world reachable from world $x$ by a measurement.
\item Invariance with respect to classical worlds for measurement: rules $\PMesubone$ and $\PMesubtwo$.
\end{itemize}
Derivations and proofs in $\MSpQR$ are defined as for $\MSQR$.
For instance, in addition to the formulas for $\uniB$ already listed for $\MSQR$, the following labelled formula schemata are all provable in $\MSpQR$ (as shown, e.g., for formula~\ref{example-three} in 
Figure~\ref{fig:example-MSpQR}):
\begin{enumerate}
\item $x: \boxdot A \varimp \Diamonddot A$ \\
(it is always possible to perform a measurement of a quantum register).
\item $x: \boxdot A \varimp \boxdot\boxdot A$ \\ 
 (measurements are composable).
\item\label{example-three} $x: \Diamonddot (A \varimp \boxdot A)$, 
i.e.~$x: \neg \boxdot \neg (A \varimp \boxdot A)$ \\ 
(it is always possible to perform a measurement with a complete reduction of a quantum register to a classical one).
\end{enumerate}
\end{itemize}

The semantics is also obtained by simple changes with respect to the definitions of 
Section~\ref{sec:semantics}.
A \emph{frame} is a tuple $\mathscr{F} = \langle W, U, P\rangle$, where 
$P \subseteq W \times W$ and $v P w$ means that $w$ is obtained by means of a measurement of $v$, with the following properties:
\begin{en}
\item $\forall v, w.\ v P w \Longrightarrow v U w$ \\
(as for (i) in Section~\ref{sec:semantics}).
\item $\forall v, w',w''.\ v P w' \ \& \ w' P w''  \Longrightarrow v P w''$ \\
 (measurements are composable).
\item $\forall v.\ \exists w.\ v P w \ \& \ w P w$ \\
(each quantum register $v$ can be reduced to a classical one $w$ by means of a measurement).
\item $\forall v, w.\ v P v \ \& \ v P w \Longrightarrow v=w$ \\
(each measurement of a classical register $v$ has $v$ as outcome).
\end{en}
\emph{Models} and \emph{structures} are defined as before, with $\I(\PMe)=P$,
while the \emph{truth} relation now comprises the clauses
\begin{center}
\begin{tabular}{lll}
$\mathscr{M},w\vDash \boxdot A$ & iff & 
$\forall w'.\ w P w' \Longrightarrow \mathscr{M}, w'\vDash A$ \\
$\mathscr{M},\I\vDash x \PMe y$ & iff & $\I(x) P \I(y)$
\end{tabular}
\end{center}
Finally, $\MSpQR$ is also sound and complete.
\begin{theorem}[Soundness and completeness of $\MSpQR$]\label{theorem:soundness-completeness-MSpQR} 
$\Gamma \vdash \alpha$ iff $\Gamma \vDash \alpha$.
\end{theorem}

\Section{Conclusions and future work}
\label{sec:conclusions}

We have shown that our modal natural deduction systems $\MSQR$ and $\MSpQR$ 
provide suitable representations of quantum register transformations. As
future work, we plan to investigate the proof theory of our systems
(e.g.~normalization, subformula property, (un)decidability), in view
of a possible mechanization of reasoning in $\MSQR$ and $\MSpQR$ (e.g.~encoding 
them into a logical framework~\cite{Pfenning01}).  We are also working at extending 
our approach to represent and reason about further quantum notions, such as
entanglement.

\bibliography{modalquantumregisters,biblio}

\begin{thebibliography}{10}

\bibitem{AbDu06}
S.~Abramsky and R.~Duncan.
\newblock A categorical quantum logic.
\newblock {\em Math. Structures Comput. Sci.}, 16(3):469--489, 2006.

\bibitem{BaltSmets04}
A.~Baltag and S.~Smets.
\newblock The logic of quantum programs.
\newblock In {\em Proceedings of the 2nd International Workshop on Quantum
  Programming Languages QPL}, 2004.

\bibitem{BalSmets06}
A.~Baltag and S.~Smets.
\newblock L{QP}: the dynamic logic of quantum information.
\newblock {\em Math. Structures Comput. Sci.}, 16(3):491--525, 2006.

\bibitem{BasDa05short}
J.-L. Basdevant and J.~Dalibard.
\newblock {\em Quantum mechanics}.
\newblock Springer-Verlag, 2005.

\bibitem{vN36}
G.~Birkhoff and J.~von Neumann.
\newblock The logic of quantum mechanics.
\newblock {\em Ann. of Math. (2)}, 37(4):823--843, 1936.

\bibitem{Chellas80}
B.~F. Chellas.
\newblock {\em Modal Logic}.
\newblock Cambridge University Press, 1980.

\bibitem{DallaCh77}
M.~L. Dalla~Chiara.
\newblock Quantum logic and physical modalities.
\newblock {\em J. Philos. Logic}, 6(4):391--404, 1977.
\newblock Special issue: Symposium on Quantum Logic (Bad Homburg, 1976).

\bibitem{DallaChiara86}
M.~L. {Dalla Chiara}.
\newblock Quantum logic.
\newblock In D.~Gabbay and F.~Guenthner, editors, {\em Handbook of
  Philosophical Logic: Volume III: Alternatives to Classical Logic}, pages
  427--469. Reidel, 1986.

\bibitem{Gabbay96}
D.~M. Gabbay.
\newblock {\em Labelled Deductive Systems}, volume~1.
\newblock Clarendon Press, 1996.

\bibitem{Mittel79}
P.~Mittelstaedt.
\newblock The modal logic of quantum logic.
\newblock {\em J. Philos. Logic}, 8(4):479--504, 1979.

\bibitem{Pfenning01}
F.~Pfenning.
\newblock Logical frameworks.
\newblock In A.~Robinson and A.~Voronkov, editors, {\em Handbook of Automated
  Reasoning}, chapter~17, pages 1063--1147. Elsevier Science and MIT Press,
  2001.

\bibitem{Prawitz65}
D.~Prawitz.
\newblock {\em Natural deduction, a proof-theoretical study}.
\newblock Almqvist and Wiksell, 1965.

\bibitem{Simpson93}
A.~Simpson.
\newblock {\em The proof theory and semantics of intuitionistic modal logic}.
\newblock PhD thesis, University of Edinburgh, UK, 1993.

\bibitem{TroelstraSchwichtenberg96}
A.~S. Troelstra and H.~Schwichtenberg.
\newblock {\em Basic proof theory}.
\newblock Cambridge University Press, 1996.

\bibitem{Vigano00a}
L.~Vigan{\`o}.
\newblock {\em {Labelled Non-Classical Logics}}.
\newblock Kluwer Academic Publishers, 2000.

\end{thebibliography}
\bibliographystyle{abbrv}

\newpage

\appendix
\section{Proof of soundness and completeness}

Theorem~\ref{theorem:soundness-completeness} follows from Theorems~\ref{theorem:soundness-MSQR}
and~\ref{theorem:completeness-MSQR} below.

\begin{theorem}[Soundness of $\MSQR$]\label{theorem:soundness-MSQR}
$\Gamma \vdash \alpha$ implies $\Gamma \vDash \alpha$.
\end{theorem}

\begin{proof}
We let $\mathscr{M}$ be an arbitrary model and prove that if $\Gamma \vdash \alpha$ then 
$\vDash \I(\Gamma)$ implies  $\vDash \I(\alpha)$ for any $\I$.
The proof proceeds by induction on the structure of the derivation of $\alpha$ from $\Gamma$.  The base case, where $\alpha \in \Gamma $, is trivial.  There is one step case for each  
rule of $\MSQR$.

Consider an application of the rule $\RAA$,
\begin{displaymath}
  \vcenter{
    \infer[\RAA]{x:A}
     {
     \infer*{y: \bot}{[x: \neg A]}
     }
    }
\end{displaymath}
where $\Gamma' \vdash y:\bot$ with $\Gamma' = \Gamma \cup \{x: \neg A\}$.  
By the induction hypothesis, $\Gamma' \vdash y: \bot$ implies $\I(\Gamma') \vDash \I(y): \bot$ 
for any $\I$.
We assume $\vDash \I(\Gamma)$ and prove $\vDash \I(x):A$. Since $\nvDash w: \bot$ for 
any world $w$, from the induction hypothesis we obtain $\nvDash \I(\Gamma')$, and thus 
$\nvDash \I(x): \neg A$, i.e.~$\vDash \I(x):A$ and $\nvDash \I(x): \bot$.

Consider an application of the rule $\bot E$,
\begin{displaymath}
  \vcenter{
    \infer[\bot E]{\alpha}{x: \bot}
    }
\end{displaymath}
with $\Gamma \vdash x:\bot$.
By the induction hypothesis, $\Gamma \vdash x: \bot$ implies $\I(\Gamma) \vDash \I(x): \bot$
for any $\I$. We assume $\vDash \I(\Gamma)$ and prove 
$\vDash \I(\alpha)$ for an arbitrary formula $\alpha$. If $\vDash \I(\Gamma)$ then 
$\vDash \I(x):\bot$ by the induction hypothesis. But since $\nvDash w: \bot$ for any world $w$, 
then $\nvDash \I(\Gamma)$ and thus $\vDash \I(\alpha)$ for any $\alpha$.

Consider an application of the rule $\bigstar I$
\begin{displaymath}
  \vcenter{
    \infer[\bigstar I]{x: \bigstar A}
     {
     \infer*{y:A}{[x Ry]}
     }
    }
\end{displaymath}
where $\Gamma'  \vdash y:A$ with $y$ fresh and
with $\Gamma' = \Gamma \cup \{x R y\}$.  By the induction hypothesis, for all interpretations
$\I$, if $\vDash \I(\Gamma)$ then $\vDash \I(y):A$. We let
$\I$ be any interpretation such that $\vDash \I(\Gamma)$, and 
show that $\vDash \I(x): \bigstar A$. Let $w$ be any world such that $\I(x) \I(R) w$ 
where $\I(R) \in \{U,M\}$ depending on $\bigstar$.  Since $\I$ can be
trivially extended to another interpretation (still called $\I$ for simplicity) by setting $\I(y)=w$,
the induction hypothesis yields $\vDash \I(y):A$, i.e.~$\vDash w:A$, and thus 
$\vDash \I(x): \bigstar A$.
 
Consider an application of the rule $\bigstar E$
\begin{displaymath}
\infer[\bigstar E]{y:A}
 {
 x: \bigstar A
 & 
 x R y
 }
\end{displaymath}
with $\Gamma_1 \vdash x: \bigstar A$ and $\Gamma_2 \vdash x R y$, and 
$\Gamma \supseteq \Gamma_1 \cup \Gamma_2$.  We assume $\vDash \I(\Gamma)$ and prove 
$\vDash \I(y):A$. By the induction hypothesis, for all interpretations
$\I$, if $\vDash \I(\Gamma_1)$ then $\vDash \I(x): \bigstar A$ and if
$\vDash \I(\Gamma_2)$ then $\vDash \I(x) \I(R) \I(y)$, where $\I(R) \in \{U,M\}$ depending on $\bigstar$.  If $\vDash \I(\Gamma)$, then $\vDash \I(x): \bigstar A$
and $\vDash \I(x) \I(R) \I(y)$, and thus $\vDash \I(y):A$.

The rules $\Unr$, $\Uns$, and $\Unt$ are sound by the properties of $U$.

The rule $\Un I$ is sound by property (i) in Definition~\ref{def:semantics}.

Consider an application of the rule $\Meser$
\begin{displaymath}
\infer[\Meser]{\alpha}{\infer*{\alpha}{[x \Me y]}}
\end{displaymath}
with $\Gamma' = \Gamma \cup \{x \Me y\}$, for $y$ fresh. 
By the induction hypothesis, $\Gamma'  \vdash \alpha$ implies  
$\I(\Gamma')  \vDash \I(\alpha)$ for any $\I$.  Let us suppose that there is an $\I'$ such 
that $\vDash \I'(\Gamma')$ and $\nvDash \I'(\alpha)$. Let us consider an $\I''$ such that
$\I''(z) = \I'(z)$ for all $z$ such that $z \neq y$ and $\I''(y)$ is the world $w$ such that $\I''(y) M w$, which exists by property (ii) in Definition~\ref{def:semantics}. Since $y$ does not occur in $\Gamma$ nor
in $\alpha$, we then have that $\vDash \I''(\Gamma')$ and $\nvDash \I''(\alpha)$, contradicting
the universality of the consequence  of the induction hypothesis. Hence, $\Meser$ is sound.

The rule $\Mesrefl$ is sound by property (iii) in Definition~\ref{def:semantics}.

Consider an application of the rule $\Mesubone$
\begin{displaymath}
\infer[\Mesubone]{\alpha(y/x)}
 {
\alpha(x)
 & 
x \Me x
 & 
x\Me y
 }
\end{displaymath}
with
$\Gamma_1 \vdash \alpha(x)$, $\Gamma_2 \vdash x \Me x$, $\Gamma_3 \vdash x \Me y$, and
$\Gamma \supseteq \Gamma_1 \cup \Gamma_2 \cup \Gamma_3$. 
We assume $\vDash \I(\Gamma)$ and prove $\vDash \I(\alpha(y/x))$.
By the induction hypothesis,
$\Gamma_1 \vdash \alpha(x)$ implies $\I(\Gamma_1) \vDash \I(\alpha(x))$,
$\Gamma_2 \vdash x \Me x$ implies $\I(\Gamma_2) \vDash \I(x) M \I(x)$, and
$\Gamma_3 \vdash x \Me y$ implies $\I(\Gamma_3) \vDash \I(x) M \I(y)$. By 
property \emph{(iv)} in Definition~\ref{def:semantics}, we then have $\I(x) = \I(y)$ and thus 
$\vDash \I(\alpha(y/x)):A$. The case for rule $\Mesubtwo$ follows analogously.
\end{proof}

To prove completeness (Theorem~\ref{theorem:completeness-MSQR}), we give some preliminary definitions and results. For simplicity, we will split each set of formulas $\Gamma$ into a pair $(\LF,\RF)$ of the subsets of l--formulas and r--formulas of $\Gamma$, and then prove $(\LF,\RF) \vDash \alpha$ implies 
$(\LF,\RF) \vdash \alpha$. We call $(\LF,\RF)$ a \emph{context} and, slightly abusing notation, 
we write $\alpha \in (\LF,\RF)$ whenever $\alpha \in \LF$ or $\alpha \in \RF$, and write 
$x \in (\LF,\RF)$ whenever the label $x$ occurs in some $\alpha \in (\LF,\RF)$.
We say that a context $(\LF,\RF)$ is \emph{consistent} iff 
$(\LF,\RF) \nvdash x:\bot$ for every $x$, so that we have:
\begin{fact}\label{fact-cons}
  If $(\LF,\RF)$ is consistent, then for every $x$ and every $A$, either $(\LF \cup \{x:A\},\RF)$ 
  is consistent or $(\LF \cup \{x: \neg A\},\RF)$ is consistent.
\end{fact}
Let $\overline{(\LF,\RF)}$ be the \emph{deductive closure} of $(\LF,\RF)$ for r--formulas under the 
rules of $\MSQR$, i.e.
\begin{displaymath}
\overline{(\LF,\RF)} \equiv \{x R y \mid (\LF,\RF)  \vdash x R y\} 
\end{displaymath}
for $R \in \{\Un,\Me\}$. We say that a context $(\LF,\RF)$ is \emph{maximally consistent} iff
\begin{enumerate}
\item it is consistent,
\item it is deductively closed for r--formulas, i.e.~$(\LF,\RF) = \overline{(\LF,\RF)}$,  and
\item for every $x$ and every $A$, either $x:A \in (\LF,\RF) $ or $x: \neg A \in (\LF,\RF) $.
\end{enumerate}

Let us write $(\LF,\RF) \vDash^{\mathscr{S}^c} \alpha$ when 
$\mathscr{S}^c \vDash (\LF,\RF)$ implies $\mathscr{S}^c \vDash \alpha$. 
Completeness follows by a Henkin--style proof, where a canonical structure
\begin{displaymath}
\mathscr{S}^c = \langle \mathscr{M}^c, \I^c \rangle = 
\langle W^c, U^c, M^c, V^c, \I^c \rangle
\end{displaymath}
is built to show that $(\LF,\RF) \nvdash \alpha$ implies 
$(\LF,\RF) \nvDash^{\mathscr{S}^c} \alpha$, 
i.e.~$\mathscr{S}^c \vDash (\LF,\RF)$ and $\mathscr{S}^c \nvDash \alpha$.

In standard proofs for unlabelled modal logics (e.g.~\cite{Chellas80}) and for other non-classical 
logics, the set $W^c$ is obtained by progressively building maximally
consistent sets of formulas, where consistency is locally checked
within each set.  In our case, given the presence of l--formulas and r--formulas, we modify the Lindenbaum lemma to extend $(\LF,\RF)$ to one single maximally consistent context 
$(\LF^*,\RF^*)$, where consistency is ``globally'' checked also against the additional
assumptions in $\RF$.\footnote{We consider only consistent contexts.  If $(\LF,\RF)$ is 
inconsistent, then $\LF, \RF \vdash x:A$ for all $x:A$, and thus completeness immediately holds 
for l--formulas.  Our language does not allow us to define inconsistency for a set of r--formulas, but, whenever $(\LF,\RF)$ is inconsistent, the canonical model built in the following is nonetheless a counter-model to non-derivable r--formulas.} The elements of $W^c$ are then built by partitioning $\LF^*$ and $\RF^*$ with respect to the labels, and the relations $R$ between the worlds are defined by exploiting the information in $\RF^*$.  

In the Lindenbaum lemma for predicate logic, a maximally consistent and $\omega$-complete 
set of formulas is inductively built by adding for every formula $\neg \forall x. A$ a 
\emph{witness} to its truth, namely a formula $\neg A[c/x]$ for some new individual constant 
$c$. This ensures that the resulting set is $\omega$-complete, i.e.~that if, for every closed 
term $t$, $A[t/x]$ is contained in the set, then so is $\forall x. A$. A similar procedure applies 
here in the case of l--formulas of the form $x: \neg \bigstar A$.  That is, together with $x: \neg
\bigstar A$ we consistently add $y: \neg A$ and $x R y$ for some new $y$, which acts as a 
\emph{witness world} to the truth of $x: \neg \bigstar A$.  This ensures that the maximally
consistent context $(\LF^*,\RF^*)$ is such that if $x R z \in (\LF^*,\RF^*)$ implies 
$z:B \in (\LF^*,\RF^*)$ for every $z$, then $x: \bigstar B \in (\LF^*,\RF^*)$, as shown in
Lemma~\ref{lemma-cs} below.  Note that in the standard completeness proof for unlabelled 
modal logics, one instead considers a canonical model $\mathscr{M}^c$ and shows that
if $w \in W^c$ and $\mathscr{M}^c,w \vDash \neg \bigstar A$, then $W^c$ also
contains a world $w'$ accessible from $w$ that serves as a witness
world to the truth of $\neg \bigstar A$ at $w$, i.e.~$\mathscr{M}^c,w' \vDash \neg A$.
\begin{lemma}\label{lemma-lindenbaum}
  Every consistent context $(\LF,\RF)$ can be extended to
  a maximally consistent context $(\LF^*,\RF^*)$.
\end{lemma}
\begin{proof}
  We first extend the language of $\MSQR$ with infinitely many new
  constants for witness worlds.  Systematically let $b$ range over
  labels, $c$ range over the new constants for witness worlds, and $a$
  range over both.  All these may be subscripted.  Let $l_1,l_2,
  \ldots$ be an enumeration of all l--formulas in the extended language;
  when $l_i$ is $a:A$, we write $\neg l_i$ for $a:\neg A$.  Starting
  from $(\LF_0,\RF_0) = (\LF,\RF)$, we inductively build a
  sequence of consistent contexts by defining
  $(\LF_{i+1},\RF_{i+1})$ to be:
  \begin{itemize}
  \item $(\LF_i,\RF_i)$, if $(\LF_i \cup \{l_{i+1}\},\RF_i)$ is inconsistent; else
  \item $(\LF_i \cup \{l_{i+1}\},\RF_i)$, if $l_{i+1}$ is not
    $a: \neg \bigstar A$; else
  \item $(\LF_i \cup \{a: \neg \bigstar A, c: \neg A\},\RF_i
    \cup\{a R c\})$ for a $c \not\in (\LF_i \cup \{a: \neg \bigstar A\},\RF_i)$, if $l_{i+1}$ is 
    $a: \neg \bigstar A$.
  \end{itemize}
  Every $(\LF_i,\RF_i)$ is consistent.  To show this we show
  that if $(\LF_i \cup \{a: \neg \bigstar A\},\RF_i)$ is
  consistent, then so is $(\LF_i \cup \{a: \neg \bigstar A, c: \neg
  A\},\RF_i \cup\{a R c\})$ for a $c \not\in (\LF_i \cup
  \{a: \neg \bigstar A\},\RF_i)$; the other cases follow by
  construction.  We proceed by contraposition.  Suppose that
  \begin{displaymath}
  (\LF_i \cup \{a: \neg \bigstar A, c: \neg A\},\RF_i \cup \{a R c\}) \vdash a_j:\bot
  \end{displaymath}
  where $c \not\in (\LF_i \cup \{a: \neg \bigstar A\},\RF_i)$.
  Then, by $\RAA$,
  \begin{displaymath}
  (\LF_i \cup \{a: \neg \bigstar A\},\RF_i \cup \{a R c\}) \vdash c: A\,,
  \end{displaymath}
  and $\bigstar I$ yields
  \begin{displaymath}
  (\LF_i \cup \{a: \neg \bigstar A\},\RF_i) \vdash a: \bigstar A\,. 
  \end{displaymath}
  Since also
  \begin{displaymath}
  (\LF_i \cup \{a: \neg \bigstar A\},\RF_i) \vdash a: \neg \bigstar A\,, 
  \end{displaymath}
  by $\neg E$ we have
  \begin{displaymath}
  (\LF_i \cup \{a: \neg \bigstar A\},\RF_i) \vdash a:\bot\,,
  \end{displaymath}
  i.e.~$(\LF_i \cup \{a: \neg \bigstar A\},\RF_i)$ is inconsistent.
  Contradiction.
  
  Now define
  \begin{displaymath}
  (\LF^*,\RF^*) = \overline{(\bigcup_{i \geq 0} \LF_i, \bigcup_{i \geq 0} \RF_i)}
  \end{displaymath}
  We show that $(\LF^*,\RF^*)$ is maximally consistent, by showing that it satisfies the three 
  conditions in the definition of maximal consistency.  For the first condition, note that if
  \begin{displaymath}
  (\bigcup_{i \geq 0} \LF_i, \bigcup_{i \geq 0} \RF_i)
  \end{displaymath}
  is consistent, then so is
  \begin{displaymath}
  \overline{(\bigcup_{i \geq 0} \LF_i, \bigcup_{i \geq 0} \RF_i)}\,.
  \end{displaymath}
  Now suppose that $(\LF^*,\RF^*)$ is inconsistent.  Then for
  some finite $(\LF',\RF')$ included in $(\LF^*,\RF^*)$
  there exists an $a$ such that $(\LF',\RF') \vdash a:\bot$.
  Every l--formula $l \in (\LF',\RF')$ is in some
  $(\LF_j,\RF_j)$.  For each $l \in (\LF',\RF')$, let
  $i_l$ be the least $j$ such that $l \in (\LF_j,\RF_j)$, and
  let $i = \mbox{max}\{i_l \mid l \in (\LF',\RF')\}$.  Then
  $(\LF',\RF') \subseteq (\LF_i,\RF_i)$, and
  $(\LF_i,\RF_i)$ is inconsistent, which is not the case.
  
  The second condition is satisfied by definition of $(\LF^*,\RF^*)$.  
  
  For the third condition, suppose that $l_{i+1} \not\in (\LF^*,\RF^*)$.  Then 
  $l_{i+1} \not\in (\LF_{i+1},\RF_{i+1})$ and $(\LF_i \cup \{l_{i+1}\},\RF_i)$ 
  is inconsistent.  Thus, by Fact~\ref{fact-cons}, $(\LF_i \cup \{\neg l_{i+1}\},\RF_i)$
  is consistent, and $\neg l_{i+1}$ is consistently added to some $(\LF_j,\RF_j)$ 
  during the construction, and therefore $\neg l_{i+1} \in (\LF^*,\RF^*)$. 
\end{proof}

The following lemma states some properties of maximally consistent
contexts.
\begin{lemma}\label{lemma-cs}
  Let $(\LF^*,\RF^*)$ be a maximally consistent context. Then
  \begin{enumerate}
  \item $(\LF^*,\RF^*) \vdash a_i R a_j\ $ iff $a_i R a_j \in (\LF^*,\RF^*)$.
  \item $(\LF^*,\RF^*) \vdash u:A\ $ iff $a:A \in (\LF^*,\RF^*)$.
  \item $a:B \varimp C \in (\LF^*,\RF^*)$ iff $a:B \in (\LF^*,\RF^*)$ implies $a:C \in (\LF^*,\RF^*)$.
  \item $a_i: \bigstar B \in (\LF^*,\RF^*)$ iff $\ a_i R a_j \in (\LF^*,\RF^*)$
    implies $a_j:B \in (\LF^*,\RF^*)$ for all $a_j$.
  \end{enumerate}
\end{lemma}
\begin{proof}
  \emph{1} and \emph{2} follow immediately by definition. We only treat \emph{4} as \emph{3} 
  follows analogously.  For the left-to-right direction, suppose that $a_i: \bigstar B \in (\LF^*,\RF^*)$.  
  Then, by (ii), $(\LF^*,\RF^*) \vdash a_i: \bigstar B$, and, by $\bigstar E$, we have 
  $(\LF^*,\RF^*) \vdash a_i R a_j$ implies $(\LF^*,\RF^*) \vdash a_j:B$ for all $a_j$.  By \emph{1} 
  and \emph{2}, conclude $a_i R a_j \in (\LF^*,\RF^*)$ implies $a_j:B \in (\LF^*,\RF^*)$ for all 
  $a_j$.  For the converse, suppose that $a_i: \bigstar B \not\in (\LF^*,\RF^*)$.  Then 
  $a_i: \neg \bigstar B \in (\LF^*,\RF^*)$, and, by the construction of $(\LF^*,\RF^*)$, there exists 
  an $a_j$ such that $a_i R a_j \in (\LF^*,\RF^*)$ and $a_j:B \not\in (\LF^*,\RF^*)$.
\end{proof}

We can now define the canonical structure 
\begin{displaymath}
\mathscr{S}^c = \langle \mathscr{M}^c, \I^c \rangle = 
\langle W^c, U^c, M^c, V^c, \I^c \rangle
\end{displaymath}
\begin{definition}
  Given a maximal consistent context $(\LF^*,\RF^*)$, we
  define the \emph{canonical structure} $\mathscr{S}^c$ as follows:
  \begin{itemize}
  \item $W^c = \{a \mid a \in (\LF^*,\RF^*)\}$,
  \item $(a_i,a_j) \in U^c$ iff $a_i \Un a_j \in (\LF^*,\RF^*)$,
  \item $(a_i,a_j) \in M^c$ iff $a_i \Me a_j \in (\LF^*,\RF^*)$,
  \item $V^c(r) = a$ iff $a:r \in (\LF^*,\RF^*)$,
  \item $\I^c(a) = a$.
  \end{itemize}
\end{definition}

Note that the standard definition of $R^c$ adopted for unlabelled modal logics, i.e.
\begin{displaymath}
  (a_i,a_j) \in R^c  \text{ iff } \{A \mid \Box A \in a_i\} \subseteq a_j\,, 
\end{displaymath}
is not applicable in our setting, since $\{A \mid \Box A \in a_i\} \subseteq a_j$ does \emph{not} 
imply $\vdash a_i R a_j$.  We would therefore be unable to prove completeness for r--formulas, 
since there would be cases, e.g.~when $\RF = \{\}$, where $\nvdash a_i R a_j$ but 
$(a_i,a_j) \in R^c$ and thus $\mathscr{S}^c \vDash a_i R a_j$.  Hence, we instead define 
$(a_i,a_j) \in R^c$ iff $a_i R a_j \in (\LF^*,\RF^*)$; note that therefore $a_i R a_j \in (\LF^*,\RF^*)$
implies $\{A \mid \Box A \in a_i\} \subseteq a_j$.  As a further comparison with the standard 
definition, note that in the canonical model the label $a$ can be identified with the set of
formulas $\{A \mid a:A \in (\LF^*,\RF^*)\}$. Moreover, we immediately have:
\begin{fact}\label{fact-R}
  $a_i R a_j \in (\LF^*,\RF^*)$ iff $(\LF^*,\RF^*) \vDash^{\mathscr{S}^c} a_i R a_j$.
\end{fact}

The deductive closure of $(\LF^*,\RF^*)$ for r--formulas ensures not only completeness for
r--formula, as shown in Theorem~\ref{theorem:completeness-MSQR} below, but also that
the conditions on $R^c$ are satisfied, so that $\mathscr{S}^c$ is really a structure
for $\MSQR$. More concretely:
\begin{itemize}
\item $U^c$ is an equivalence relation by construction and rules $\Unr$, $\Uns$, and $\Unt$. 
For instance, for transitivity, consider an arbitrary context $(\LF,\RF)$ from which we build 
$\mathscr{S}^c$. Assume $(a_i,a_j) \in U^c$ and $(a_j,a_k) \in U^c$.  Then 
$a_i \Un a_j \in (\LF^*,\RF^*)$ and $a_j \Un a_k \in (\LF^*,\RF^*)$. Since $(\LF^*,\RF^*)$ is
deductively closed, by~\emph{1} in Lemma~\ref{lemma-cs} and rule $\Unt$, we have 
$a_i \Un a_k \in (\LF^*,\RF^*)$. Thus, $(a_i,u_k) \in U^c$ and $U^c$ is indeed transitive.

\item $\forall v, w \in W^c.\ v M w \Longrightarrow v U w$ holds by construction and rule $\Un I$.

\item $\forall v \in W^c.\ \exists w \in W^c.\ v M w$ holds by construction and rule $\Meser$.
For the sake of contradiction, consider an arbitrary $a_i$ and a variable $a_j'$ 
that do not satisfy the property. Define $(\LF',\RF') = (\LF^*,\RF^*) \cup \{a_i \Me a_j'\}$. 
Then it cannot be the case that $(\LF',\RF') \vdash \alpha$, for otherwise $(\LF^*,
\RF^*) \vdash \alpha$ would be derivable by an application of the rule $\Meser$. Thus, 
$(\LF',\RF') \nvdash \alpha$. But then $(\LF',\RF')$ must be in the chain of contexts built in 
Lemma~\ref{lemma-cs}. So, by the maximality of $(\LF^*,\RF^*)$, we have that 
$(\LF',\RF') = (\LF^*,\RF^*)$, contradicting our assumption. Hence, for some $a_j$, the r--formula 
$a_i \Me a_j$ is in $(\LF^*,\RF^*)$, which is what we had to show.

\item $\forall v, w \in W^c.\ v M w \Longrightarrow w M w$ holds by construction and rule $\Mesrefl$.

\item $\forall v, w \in W^c.\ v M v \ \& \ v M w \Longrightarrow v=w$  holds by construction and
rules $\Mesubone$ and $\Mesubtwo$ since $v$ is a classical world.
Consider an arbitrary context $(\LF,\RF)$ from which we build $\mathscr{S}^c$ and assume
$(a_i,a_i) \in M^c$ and $(a_i,a_j) \in M^c$. Then $a_i \Me a_i \in (\LF^*,\RF^*)$ and 
$a_i \Me a_j \in (\LF^*,\RF^*)$. Thus, for each $a_i:A \in (\LF^*,\RF^*)$, we also have 
$a_j:A \in (\LF^*,\RF^*)$; otherwise, since $(\LF^*,\RF^*)$ is deductively closed, we would have 
$a_j: \neg A \in (\LF^*,\RF^*)$ and also $a_j: A \in (\LF^*,\RF^*)$ by \emph{1} in 
Lemma~\ref{lemma-cs} and rule $\Mesubone$, and thus a contradiction. Similarly, if 
$a_j:A \in (\LF^*,\RF^*)$ then $a_i:A \in (\LF^*,\RF^*)$ by rule $\Mesubtwo$. Hence, for each 
m--formula $A$, we have that $a_i:A \in (\LF^*,\RF^*)$ iff $a_j:A \in (\LF^*,\RF^*)$, which
means that $a_i$ and $a_j$ are equal with respect to m--formulas. 

Under the same assumptions, we can similarly show that $a_i$ and $a_j$ are equal with 
respect to r--formulas, i.e.~that whenever $(\LF^*,\RF^*)$ contains an r--formula that includes $
a_i$ then it also contains the same r--formula with $a_j$ substituted for $a_i$, and vice versa. 
To this end, we must consider 8 different cases corresponding to 8 different r--formulas. 
\begin{itemize}
\item If $a_k \Un a_i \in (\LF^*,\RF^*)$ for some $a_k$, then from the assumption that 
$a_i \Me a_j \in (\LF^*,\RF^*)$ we have $a_i \Un a_j \in (\LF^*,\RF^*)$, 
by \emph{1} in Lemma~\ref{lemma-cs} and rule $\Un I$. Therefore, 
$a_k \Un a_j \in (\LF^*,\RF^*)$ by rule $\Unt$.  
\item We can reason similarly for $a_j \Un a_k \in (\LF^*,\RF^*)$ and also apply rules $\Un I$ 
and $\Unt$ to conclude that then also $a_i \Un a_k \in (\LF^*,\RF^*)$. 
\item If $a_i \Un a_k \in (\LF^*,\RF^*)$ for some $a_k$, then from the assumption that 
$a_i \Me a_j \in (\LF^*,\RF^*)$ we have $a_i \Un a_j \in (\LF^*,\RF^*)$, 
by \emph{1} in Lemma~\ref{lemma-cs} and rule $\Un I$, and thus 
$a_j \Un a_i \in (\LF^*,\RF^*)$, by rule $\Uns$. Therefore, 
$a_j \Un a_k \in (\LF^*,\RF^*)$ by rule $\Unt$.  
\item We can reason similarly for $a_k \Un a_j \in (\LF^*,\RF^*)$ and also apply rules $\Un I$,
$\Uns$, and $\Unt$ to conclude that then also $a_k \Un a_i \in (\LF^*,\RF^*)$. 
\item If $a_k \Me a_i \in (\LF^*,\RF^*)$ for some $a_k$, then from the assumption that 
$a_i \Me a_j \in (\LF^*,\RF^*)$ we have $a_k \Me a_j \in (\LF^*,\RF^*)$,
by \emph{1} in Lemma~\ref{lemma-cs} and the derived rule $\Metrans$.
\item We can reason similarly for $a_j \Me a_k \in (\LF^*,\RF^*)$ and also apply rule $\Metrans$
to conclude that then also $a_i \Un a_k \in (\LF^*,\RF^*)$. 
\item If $a_i \Me a_k \in (\LF^*,\RF^*)$ for some $a_k$, then from the assumptions that 
$a_i \Me a_i \in (\LF^*,\RF^*)$ and $a_i \Me a_j \in (\LF^*,\RF^*)$ we have 
$a_j \Me a_k \in (\LF^*,\RF^*)$, by \emph{1} in Lemma~\ref{lemma-cs} and rule $\Mesubone$.  
\item We can reason similarly for $a_k \Me a_j \in (\LF^*,\RF^*)$ and apply rule $\Mesubtwo$
to conclude that then also $a_k \Me a_i \in (\LF^*,\RF^*)$. 
\end{itemize}
Hence, $a_i$ and $a_j$ are equal also with respect to r--formulas, and thus
$a_i = a_j$ whenever $(a_i,a_i) \in M^c$ and $(a_i,a_j) \in M^c$, which is what we had to show.
\end{itemize}

By Lemma~\ref{lemma-cs} and Fact~\ref{fact-R}, it follows that:
\begin{lemma}\label{lemma-truth}
  $a:A \in (\LF^*,\RF^*)$ iff $(\LF^*,\RF^*) \vDash^{\mathscr{S}^c} a:A$.
\end{lemma}
\begin{proof}
  We proceed by induction on the grade of $a:A$, and we treat only the step case where 
  $a:A$ is $a_i:\bigstar B$; the other cases follow analogously. For the left-to-right direction, 
  assume $a_i:\bigstar B \in (\LF^*,\RF^*)$.  Then, by Lemma~\ref{lemma-cs}, 
  $a_i R a_j \in (\LF^*,\RF^*)$ implies $a_j:B \in (\LF^*,\RF^*)$, for all $a_j$.
  Fact~\ref{fact-R} and the induction hypothesis yield that 
  $(\LF^*,\RF^*) \vDash^{\mathscr{S}^c} a_j:B$ for all 
  $a_j$ such that $(\LF^*,\RF^*) \vDash^{\mathscr{S}^c} a_i \I^c(R) a_j$, 
  i.e.~$(\LF^*,\RF^*) \vDash^{\mathscr{S}^c} a_i:\bigstar B$ by Definition~\ref{def:truth}.
  For the converse, assume $a_i:\neg \bigstar B \in (\LF^*,\RF^*)$.  Then, by
  Lemma~\ref{lemma-cs}, $a_i R a_j \in (\LF^*,\RF^*)$ and $a_j:\neg B \in (\LF^*,\RF^*)$, for 
  some $a_j$. Fact~\ref{fact-R} and the induction hypothesis yield
  $(\LF^*,\RF^*) \vDash^{\mathscr{S}^c} a_i R a_j$ and 
  $(\LF^*,\RF^*) \vDash^{\mathscr{S}^c} a_j:\neg B$, 
  i.e.~$(\LF^*,\RF^*) \vDash^{\mathscr{S}^c} a_i:\neg \bigstar B$ by Definition~\ref{def:truth}.
\end{proof}

We can now finally show:
\begin{theorem}[Completeness of $\MSQR$]\label{theorem:completeness-MSQR}
$\Gamma \vDash \alpha$ implies $\Gamma \vdash \alpha$.
\end{theorem}
\begin{proof}
  If $(\LF,\RF) \nvdash b_i R b_j$, then $b_i R b_j \not\in (\LF^*,\RF^*)$, and thus 
  $(\LF^*,\RF^*) \nvDash^{\mathscr{S}^c} b_i R b_j$ by Fact~\ref{fact-R}.  
  
  If $(\LF,\RF) \nvdash b:A$, then $(\LF \cup \{b:\neg A\},\RF)$ is consistent; otherwise
  there exists a $b_i$ such that $(\LF \cup \{b: \neg A\},\RF) \vdash b_i: \bot$, and then 
  $(\LF,\RF) \vdash b:A$. Therefore, by Lemma~\ref{lemma-lindenbaum}, 
  $(\LF \cup \{b: \neg A\},\RF)$ is included in a maximally consistent
  context $((\LF \cup \{b: \neg A\})^*,\RF^*)$.  Then, by
  Lemma~\ref{lemma-truth}, $((\LF \cup \{b: \neg A\})^*,\RF^*) \vDash^{M^C} b: \neg A$, 
  i.e.~$((\LF \cup\{w: \neg A\})^*,\RF^*) \nvDash^{\mathscr{S}^c} b:A$, and thus 
  $(\LF,\RF) \nvDash^{\mathscr{S}^c} w:A$.
\end{proof}

We can reason similarly to show the soundness and completeness of $\MSpQR$ with respect to the corresponding semantics: 
Theorem~\ref{theorem:soundness-completeness-MSpQR} follows from Theorems~\ref{theorem:soundness-MSpQR}
and~\ref{theorem:completeness-MSpQR} below.

\begin{theorem}[Soundness of $\MSpQR$]\label{theorem:soundness-MSpQR}
$\Gamma \vdash \alpha$ implies $\Gamma \vDash \alpha$.
\end{theorem}

\begin{proof}
We let $\mathscr{M}$ be an arbitrary model and prove that if $\Gamma \vdash \alpha$ then 
$\vDash \I(\Gamma)$ implies  $\vDash \I(\alpha)$ for any $\I$.
The proof proceeds by induction on the structure of the derivation of $\alpha$ from $\Gamma$.  The base case, where $\alpha \in \Gamma $, is trivial.  There is one step case for each 
rule of $\MSpQR$, where the soundness of the rules $\varimp I$, $\varimp E$, $\RAA$, $\bot E$, 
$\Unr$, $\Uns$, $\Unt$ follows exactly like in the proof of Theorem~\ref{theorem:soundness-MSQR}.
 
The soundness of the rules $\bigstar I$ and $\bigstar E$ follows exactly like in the proof of 
Theorem~\ref{theorem:soundness-MSQR}, with the only difference that when $\bigstar$ is 
$\boxdot$ then $R$ is $\PMe$.

The rule $\PMe\Un I$ is sound by property \emph{(i)} in the definition of the semantics for $\MSpQR$.

The rule $\PMetrans$ is sound by property \emph{(ii)} in the definition of the semantics for $\MSpQR$.

The soundness of the rule  $\class$ follows like for the soundness
of the rule $\Meser$ in the proof of 
Theorem~\ref{theorem:soundness-MSQR}, this time exploiting property  \emph{(iii)} in the 
definition of the semantics for $\MSpQR$.

The soundness of the rules $\PMesubone$ and $\PMesubtwo$ follows like for the soundness
of the rules $\Mesubone$ and $\Mesubtwo$ in the proof of 
Theorem~\ref{theorem:soundness-MSQR}, this time exploiting property  \emph{(iv)} in the 
definition of the semantics for $\MSpQR$.
\end{proof}

To prove completeness (Theorem~\ref{theorem:completeness-MSQR}), we proceed like for the case
of $\MSQR$, mutatis mutandis in the construction of the canonical model. In particular, given a 
maximal consistent context $(\LF^*,\RF^*)$, we define the canonical structure
$\mathscr{S}^c = \langle W^c, U^c, P^c, V^c, \I^c \rangle$ by setting 
\begin{itemize}
\item $(a_i,a_j) \in P^c$ iff $a_i \PMe a_j \in (\LF^*,\RF^*)$.
\end{itemize}
To show that the conditions on $R^c$ are satisfied, so that $\mathscr{S}^c$ is really a structure
for $\MSpQR$, we reuse the results proved for $\MSQR$ and in addition show the following:
\begin{itemize}
\item $\forall v, w \in W^c.\ v P w \Longrightarrow v U w$ holds by construction and rule $\PMe\Un I$.

\item $\forall v, w', w'' \in W^c.\ v P w' \ \& \ w' P w'' \Longrightarrow v P w''$ holds by construction 
and rule $\PMetrans$.

\item $\forall v \in W^c.\ \exists w \in W^c.\ v P w \ \& \ w P w$ holds by construction and rule 
$\class$. For the sake of contradiction, consider an arbitrary $a_i$ and a variable $a_j'$ 
that do not satisfy the property. Define $(\LF',\RF') = (\LF^*,\RF^*) \cup \{a_i \PMe a_j', a_j' \PMe a_j'\}$. 
Then it cannot be the case that $(\LF',\RF') \vdash \alpha$, for otherwise $(\LF^*,
\RF^*) \vdash \alpha$ would be derivable by an application of the rule $\class$. Thus, 
$(\LF',\RF') \nvdash \alpha$. But then $(\LF',\RF')$ must be in the chain of contexts built in 
Lemma~\ref{lemma-cs}. So, by the maximality of $(\LF^*,\RF^*)$, we have that 
$(\LF',\RF') = (\LF^*,\RF^*)$, contradicting our assumption. Hence, for some $a_j$, the r--formulas 
$a_i \Me a_j$ and $a_j \Me a_j$ are both in $(\LF^*,\RF^*)$, which is what we had to show.

\item $\forall v, w \in W^c.\ v P v \ \& \ v P w \Longrightarrow v=w$  holds by construction and
rules $\PMesubone$ and $\PMesubtwo$ since $v$ is a classical world.
Consider an arbitrary context $(\LF,\RF)$ from which we build $\mathscr{S}^c$ and assume
$(a_i,a_i) \in P^c$ and $(a_i,a_j) \in P^c$. Then $a_i \PMe a_i \in (\LF^*,\RF^*)$ and 
$a_i \PMe a_j \in (\LF^*,\RF^*)$. Thus, for each $a_i:A \in (\LF^*,\RF^*)$, we also have 
$a_j:A \in (\LF^*,\RF^*)$; otherwise, since $(\LF^*,\RF^*)$ is deductively closed, we would have 
$a_j: \neg A \in (\LF^*,\RF^*)$ and also $a_j: A \in (\LF^*,\RF^*)$ by \emph{1} in 
Lemma~\ref{lemma-cs} and rule $\PMesubone$, and thus a contradiction. Similarly, if 
$a_j:A \in (\LF^*,\RF^*)$ then $a_i:A \in (\LF^*,\RF^*)$ by rule $\PMesubtwo$. Hence, for each 
m--formula $A$, we have that $a_i:A \in (\LF^*,\RF^*)$ iff $a_j:A \in (\LF^*,\RF^*)$, which
means that $a_i$ and $a_j$ are equal with respect to m--formulas. 

Under the same assumptions, we can similarly show that $a_i$ and $a_j$ are equal with 
respect to r--formulas, i.e.~that whenever $(\LF^*,\RF^*)$ contains an r--formula that includes $
a_i$ then it also contains the same r--formula with $a_j$ substituted for $a_i$, and vice versa. 
To this end, we must consider 8 different cases corresponding to 8 different r--formulas. 
\begin{itemize}
\item If $a_k \Un a_i \in (\LF^*,\RF^*)$ for some $a_k$, then from the assumption that 
$a_i \PMe a_j \in (\LF^*,\RF^*)$ we have $a_i \Un a_j \in (\LF^*,\RF^*)$, 
by \emph{1} in Lemma~\ref{lemma-cs} and rule $\PMe\Un I$. Therefore, 
$a_k \Un a_j \in (\LF^*,\RF^*)$ by rule $\Unt$.  
\item We can reason similarly for $a_j \Un a_k \in (\LF^*,\RF^*)$ and also apply rules $\PMe\Un I$ 
and $\Unt$ to conclude that then also $a_i \Un a_k \in (\LF^*,\RF^*)$. 
\item If $a_i \Un a_k \in (\LF^*,\RF^*)$ for some $a_k$, then from the assumption that 
$a_i \PMe a_j \in (\LF^*,\RF^*)$ we have $a_i \Un a_j \in (\LF^*,\RF^*)$, 
by \emph{1} in Lemma~\ref{lemma-cs} and rule $\PMe\Un I$, and thus 
$a_j \Un a_i \in (\LF^*,\RF^*)$, by rule $\Uns$. Therefore, 
$a_j \Un a_k \in (\LF^*,\RF^*)$ by rule $\Unt$.  
\item We can reason similarly for $a_k \Un a_j \in (\LF^*,\RF^*)$ and also apply rules $\PMe\Un I$,
$\Uns$, and $\Unt$ to conclude that then also $a_k \Un a_i \in (\LF^*,\RF^*)$. 
\item If $a_k \PMe a_i \in (\LF^*,\RF^*)$ for some $a_k$, then from the assumption that 
$a_i \PMe a_j \in (\LF^*,\RF^*)$ we have $a_k \PMe a_j \in (\LF^*,\RF^*)$,
by \emph{1} in Lemma~\ref{lemma-cs} and the rule $\PMetrans$.
\item We can reason similarly for $a_j \PMe a_k \in (\LF^*,\RF^*)$ and also apply rule $\PMetrans$
to conclude that then also $a_i \Un a_k \in (\LF^*,\RF^*)$. 
\item If $a_i \PMe a_k \in (\LF^*,\RF^*)$ for some $a_k$, then from the assumptions that 
$a_i \PMe a_i \in (\LF^*,\RF^*)$ and $a_i \PMe a_j \in (\LF^*,\RF^*)$ we have 
$a_j \PMe a_k \in (\LF^*,\RF^*)$, by \emph{1} in Lemma~\ref{lemma-cs} and rule $\PMesubone$.  
\item We can reason similarly for $a_k \PMe a_j \in (\LF^*,\RF^*)$ and apply rule $\PMesubtwo$
to conclude that then also $a_k \PMe a_i \in (\LF^*,\RF^*)$. 
\end{itemize}
Hence, $a_i$ and $a_j$ are equal also with respect to r--formulas, and thus
$a_i = a_j$ whenever $(a_i,a_i) \in P^c$ and $(a_i,a_j) \in P^c$, which is what we had to show.
\end{itemize}

Proceeding like for $\MSQR$, we then have:
\begin{theorem}[Completeness of $\MSpQR$]\label{theorem:completeness-MSpQR}
$\Gamma \vDash \alpha$ implies $\Gamma \vdash \alpha$. \hfill $\triangle$
\end{theorem}

\end{document}